\newcommand{\CLASSINPUTtoptextmargin}{2.2cm}
\newcommand{\CLASSINPUTbottomtextmargin}{4cm}
\documentclass[conference]{IEEEtran}
\IEEEoverridecommandlockouts
\usepackage{cite}
\usepackage{amsmath,amssymb,amsfonts}
\usepackage{algorithmic}
\usepackage{graphicx}
\usepackage{textcomp}
\usepackage{xcolor}
\def\BibTeX{{\rm B\kern-.05em{\sc i\kern-.025em b}\kern-.08em
    T\kern-.1667em\lower.7ex\hbox{E}\kern-.125emX}}

\begin{document}

\title{Federated Learning in Adversarial Environments: Testbed Design and Poisoning Resilience in Cybersecurity\vspace{-4mm}
}

% \author{
% \vspace{-3mm}
% Anonymous Authors
% \vspace{-3mm}
% }

\vspace{-4mm}
\author{
\IEEEauthorblockN{
Hao Jian Huang, 
Hakan T. Otal, and
M. Abdullah Canbaz}

\IEEEauthorblockA{
\textit{
Department of Information Science and Technology
} \\
\textit{University at Albany, SUNY, New York, USA} \\
hhuang22, hotal, mcanbaz [at] albany [dot] edu 
\vspace{-4mm}
}
}

\maketitle
\vspace{-4mm}
\begin{abstract}
This paper presents the design and implementation of a Federated Learning (FL) testbed, focusing on its application in cybersecurity and evaluating its resilience against poisoning attacks. Federated Learning allows multiple clients to collaboratively train a global model while keeping their data decentralized, addressing critical needs for data privacy and security, particularly in sensitive fields like cybersecurity. Our testbed, built using Raspberry Pi and Nvidia Jetson hardware by running the Flower framework, facilitates experimentation with various FL frameworks, assessing their performance, scalability, and ease of integration. Through a case study on federated intrusion detection systems, the testbed's capabilities are shown in detecting anomalies and securing critical infrastructure without exposing sensitive network data. Comprehensive poisoning tests, targeting both model and data integrity, evaluate the system's robustness under adversarial conditions. The results show that while federated learning enhances data privacy and distributed learning, it remains vulnerable to poisoning attacks, which must be mitigated to ensure its reliability in real-world applications. 

\end{abstract}


\vspace{-4mm}
\section{Introduction}

In today’s data-driven world, the ability to share and analyze data collaboratively is crucial, yet data privacy remains a paramount concern, particularly in sectors such as healthcare, finance, and cybersecurity, where sensitive information must be protected under stringent regulations such as GDPR and HIPAA \cite{Ganjoo_2022, Guduri_2024}. The recent Presidential Executive Order on the \textit{Safe, Secure, and Trustworthy Development and Use of AI} underscores the growing emphasis on ensuring AI systems, including FL, are both secure and privacy-preserving \cite{biden2023executive}.

FL enables decentralized model training, preserving data privacy by sharing only model updates instead of raw data \cite{Kairouz2021}. Originally introduced by Google, FL allows individual devices or organizations to train local models on their own datasets while sharing only model updates (e.g., gradients or parameters) with a central server \cite{Banabilah2022}. By maintaining data privacy and decentralization, FL mitigates the risks associated with centralized data collection and complies with data protection regulations \cite{Zhang_2021}.

The decentralized nature of FL is particularly beneficial in fields where data sharing is restricted due to legal or ethical considerations. In healthcare, hospitals can collaboratively develop predictive models for disease diagnosis without sharing sensitive patient records \cite{Xu_2021}. Similarly, in cybersecurity, organizations can detect threats like malware or phishing attacks without exposing proprietary network data \cite{Nassar2021}. Furthermore, by limiting communication to model updates rather than raw data, FL reduces the risk of data leakage during transmission \cite{jin2023federated}.

Despite these advantages, FL is not without its challenges. The decentralized nature of FL introduces new vulnerabilities, particularly in the form of poisoning attacks. In a poisoning attack, an adversary deliberately manipulates the training process by introducing malicious data or altering model updates, which can lead to wrong predictions or even the failure of the global model to converge \cite{Ganjoo_2022}. These attacks can take several forms, including data poisoning, where the attacker corrupts the training data, and model poisoning, where the attacker directly manipulates the model parameters \cite{Cina2024}. Such attacks pose significant risks in critical domains like healthcare and cybersecurity, where accuracy and integrity are vital \cite{Hu_2024}.

Recognizing these challenges, researchers have explored various defense mechanisms to protect FL systems from poisoning attacks. One prominent approach is the use of Byzantine Robust Aggregation (BRA), which aims to filter out malicious updates during the model aggregation process, ensuring that only legitimate updates are incorporated into the global model \cite{Chaalan2024} and some using methods like Krum or Median to filter malicious updates based on distance or statistics \cite{Cajaraville-Aboy_2024}. Differential Privacy (DP) adds noise to updates, providing formal privacy guarantees with inherent trade-offs analyzed theoretically in works like \cite{Wei_2019}. Another strategy involves integrating differential privacy techniques, which add noise to model updates to obscure the contribution of any single participant, thereby preventing adversaries from inferring sensitive information or manipulating the model \cite{Priya_2021}. In addition, blockchain-based approaches have been proposed to enhance the security of FL systems by creating a decentralized, immutable ledger that records all interactions between clients and the central server, ensuring transparency and trust \cite{Guduri_2024}. Other defense strategies include trust-based weighting (e.g., FLTrust), using gradient sign information (e.g., SignGuard), or proactive alarming (e.g., SIREN), which offer alternative ways to enhance robustness against malicious clients \cite{Cajaraville-Aboy_2024}.

While existing research explores FL in cybersecurity, limited real-world testbeds evaluate its vulnerabilities under targeted adversarial settings. Our testbed bridges this gap by enabling systematic poisoning attack simulations on IoT-scale devices. In this paper, we introduce a novel approach by developing a FL testbed tailored to evaluate the nuanced effects of poisoning attacks in cybersecurity contexts. We conduct a measurement study to systematically analyze the impact of both data and model poisoning on FL systems. By simulating real-world adversarial scenarios, we investigate how malicious inputs compromise model performance, convergence, and accuracy in a small-scale setting with 3-5 clients. While our testbed examines poisoning resilience in a small-scale setting, decentralized FL (DFL) architectures aim to enhance scalability compared to traditional centralized FL by removing the central server bottleneck \cite{Cajaraville-Aboy_2024}. Evaluating security mechanisms across different scales and architectures remains an important research direction. This research offers an exploration of the vulnerabilities inherent in FL, providing critical insights that can guide the design of more resilient FL architectures. Our findings contribute to a deeper understanding of how adversarial attacks can degrade system integrity, shaping future strategies for safeguarding FL in privacy-sensitive and security-critical environments.

\vspace{-2mm}
\section{Methodology}
\vspace{-2mm}

To address the challenges of poisoning attacks in FL and evaluate defense mechanisms, we developed a specialized FL testbed. The testbed is designed to replicate real-world scenarios where data and model poisoning can occur, while providing a controlled environment to assess the resilience of such systems. By leveraging this setup, we aim to understand the impacts of adversarial interventions and compare the performance of poisoned versus non-poisoned datasets, highlighting key vulnerabilities and potential mitigation strategies.

\subsection{Overview of System Architecture}
\begin{figure}[!t]
\centering
\includegraphics[width=\columnwidth]{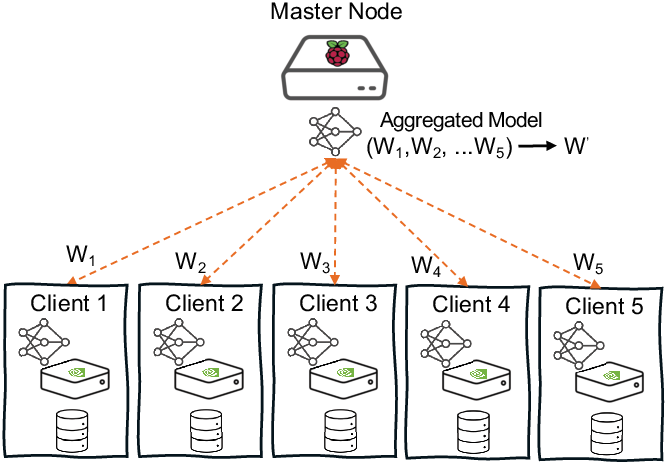}
\vspace{-4mm}
\caption{System architecture for the testbed, illustrating the key components and interactions within the FL setup.}
\label{fig:testbed}
\vspace{-3mm}
\end{figure}
The FL testbed, illustrated in Figure~\ref{fig:testbed}, follows a client-server architecture implemented using the Flower framework \cite{beutel2020flower}, which provides a flexible platform for centralized FL workflows. The setup consists of three main components: client nodes, an aggregation server, and communication protocols.

At the core of the testbed is the master node, a Raspberry Pi 4, chosen for its lightweight footprint and ability to simulate an edge server environment. The master node orchestrates the training process, initializing and distributing the global model to the client nodes, which are Nvidia Jetson Nano devices. Each client operates independently, processing local data and sending only model updates (e.g., gradients or weights) back to the master node. This decentralized approach ensures that sensitive data remains on the client devices, reflecting real-world FL scenarios where devices vary in computational power, network latency, and data quality (non-IID). The master node aggregates these updates, typically using Federated Averaging (FedAvg) \cite{9850408}, to refine the global model, which is then redistributed to the clients for further training in iterative rounds.

FedAvg was chosen as our aggregation algorithm because it is a standard baseline in FL research, known for its communication efficiency, scalability, and strong empirical performance on non-IID data distributions. In our FL testbed, clients are simulating resource-constrained IoT devices, making FedAvg particularly suitable due to its low communication overhead and adaptability to heterogeneous client datasets. 

The communication between the server and the clients is facilitated by bi-directional gRPC streams, enabling fast parallel data exchanges. Also, Flower framework and this architecture allows scalability up to 1000s of clients as shown in their paper \cite{beutel2020flower} and allowing adjustments to the number of clients based on specific experiments or use cases.

In this setup, we explore two key use cases: (i) establish a baseline performance using clean datasets and (ii) introduce data poisoning attacks at the client level. By manipulating client data in the poisoned scenario, we analyze how adversarial actions impact the global model's accuracy, convergence, and robustness. 

\subsection{Software and Hardware Setup} 
The FL testbed is built around a Raspberry Pi 4 as the central server, running Pi OS, paired with five Nvidia Jetson Nano devices acting as client nodes. Each Jetson Nano operates on Ubuntu OS, providing a lightweight yet capable environment for edge computing tasks.

For the software, Visual Studio Code as IDE and Python as the primary programming language are used. The machine learning models were implemented using TensorFlow~\cite{abadi2016tensorflow} and Scikit-learn~\cite{pedregosa2011scikit} to develop and train an intrusion detection model based on a multi-layer perceptron architecture. To support data manipulation and analysis, Numpy \cite{oliphant2006guide} and Pandas \cite{mckinney2011pandas} are utilized for data handling and preprocessing. Additionally, Matplotlib \cite{tosi2009matplotlib} was used for visualizing the results of the training processes.

\subsection{Comparison of FL Frameworks}
During the testbed development, we evaluated open-source FL frameworks based on library compatibility, scalability, privacy, and suitability for resource-constrained environments. Four key candidates emerged:

\begin{itemize}
\item \textbf{TFF}: Integrates with TensorFlow but lacks flexibility for other ML libraries \cite{10303221}.
\item \textbf{PySyft}: Strong in privacy-preserving FL but complex and resource-intensive \cite{9359017}.
\item \textbf{FATE}: Designed for large-scale enterprise FL, making it less suited for research-focused setups \cite{9153560}.
\item \textbf{Flower}: Supports multiple ML libraries, is lightweight, and scales efficiently, making it ideal for IoT and academic use \cite{beutel2020flower}.
\end{itemize}

We selected Flower for its flexibility, lightweight architecture, and strong community support. It integrates well with PyTorch and Scikit-learn, operates efficiently on devices like Nvidia Jetson Nano, and scales easily.

\vspace{-2mm}
\subsection{Datasets}
For our experiments, we utilized the DNP3 intrusion detection dataset \cite{s7h0-b081-22}, which contains labeled network traffic features extracted using CICFlowMeter~\cite{lashkari2017characterization}. The dataset includes 83 features, with preprocessing steps reducing it to 76 numerical features after encoding categorical variables and handling missing values. The dataset was split 70/30 for training and testing, with training data further partitioned among clients to simulate a realistic FL setting. Each client received a non-IID subset of the data, reflecting real-world variability in network environments. The intrusion detection task involved classifying network flows into 11 categories, including benign traffic and various cyberattacks targeting the DNP3 protocol. DNP3 is widely used in industrial control systems, making it a critical attack vector in cybersecurity. Evaluating FL on this dataset provides insights into securing industrial networks while maintaining data privacy.

\section{Experimental Results}

In this section, we present the results of our FL experiments, focusing on a cybersecurity use case involving intrusion detection within critical infrastructure systems. Specifically, our objective is to identify and mitigate malicious activities within Distributed Network Protocol 3 (DNP3) communications. The ability to detect intrusions in DNP3 traffic is vital for preventing unauthorized access, system disruptions, and potential security breaches\cite{ji2023ae}.

To address this challenge, we implemented a multi-layer perceptron (MLP) model with one hidden layer, trained on the preprocessed DNP3 intrusion detection dataset \cite{s7h0-b081-22}. During preprocessing, non-numerical values were encoded into numerical features, and any \textit{Inf} or \textit{NaN} values were removed, resulting in a final dataset of 76 features and 7,326 rows.  

For small-scale FL experiments, the dataset is relatively limited but sufficient to evaluate the feasibility of distributed intrusion detection. Its size is particularly appropriate given the computational constraints of IoT and edge devices in FL scenarios. However, in real-world deployments, larger datasets are typically required to enhance generalization and robustness.

\begin{figure*}[!htb]
\vspace{2mm}
\centering
\includegraphics[width=.95\linewidth]{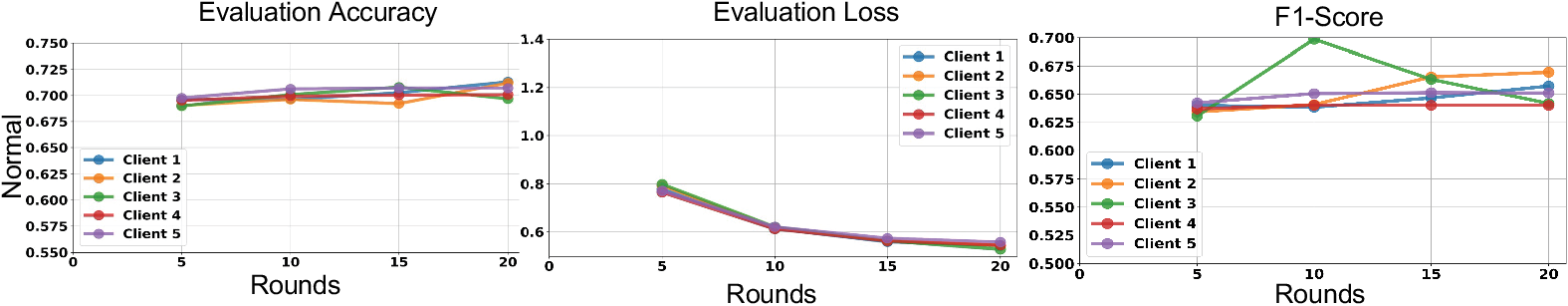}
\caption{The performance of FL testbed clients models with normal data}
\label{fig:normal-5client}
\vspace{-2mm}
\end{figure*} 

The intrusion detection system we developed is designed to classify network traffic into 11 different categories, corresponding to specific types of DNP3 traffic and attacks. Our model processes 76 input features, feeding them through a hidden layer containing 50 neurons activated by the ReLU function. To mitigate overfitting, a dropout layer deactivates 20\% of the neurons during training. The final output layer employs a Softmax activation function to classify the input into one of the 11 labels.

\begin{figure*}[!htb]
\centering
\includegraphics[width=0.95\linewidth]{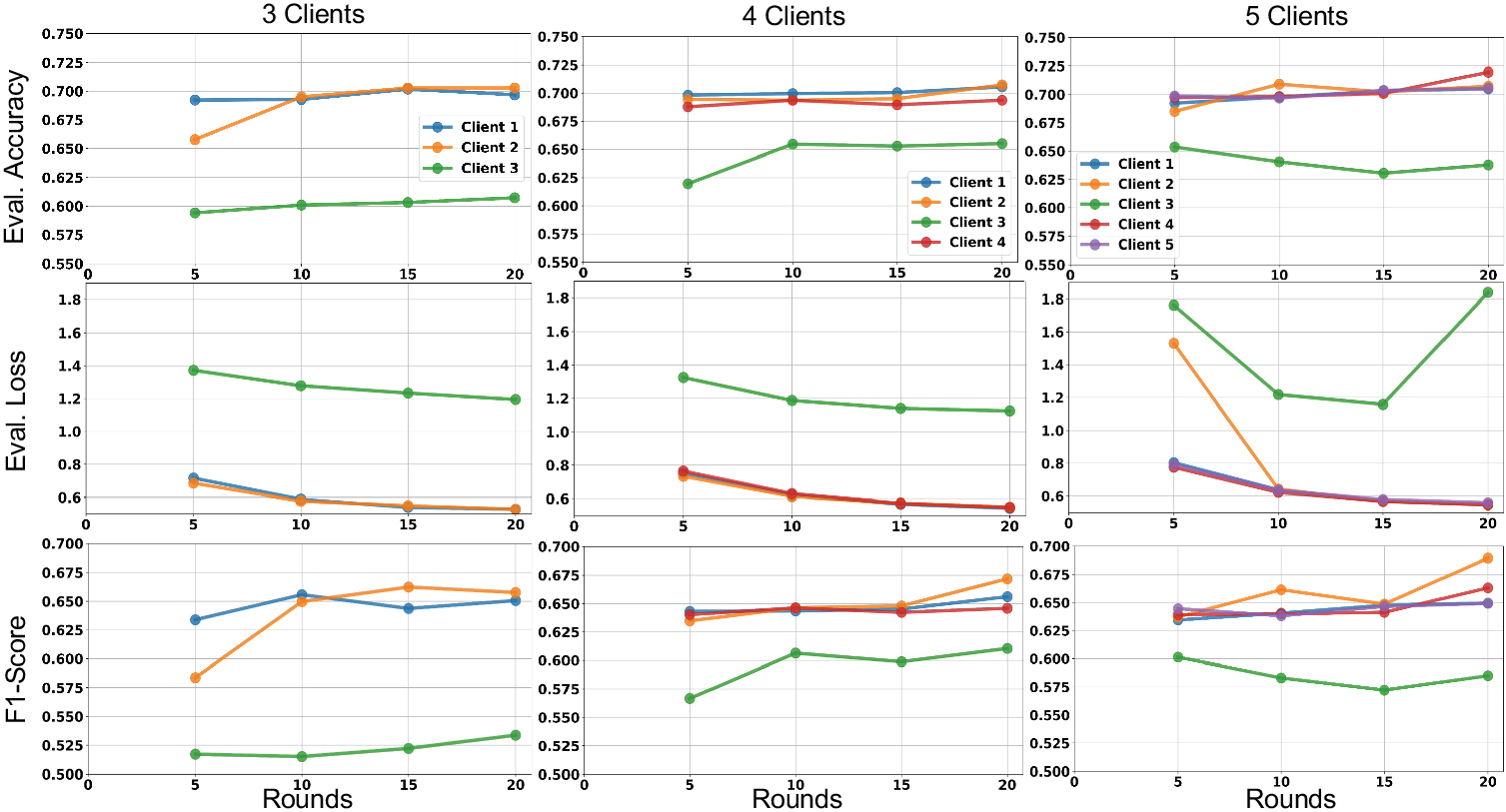}
\caption{The performance of each experiment with multiple clients (3,4,5) in evaluation accuracy, loss, and F1-Score}
\label{fig:Client-Poisoned}
%\vspace{-3mm}
\end{figure*} 

Training for our experiments was carried out over 20 rounds, with each round consisting of 20 epochs. The Adam optimizer was employed due to its efficiency in handling sparse gradients and its adaptive learning rate properties, making it well-suited for FL environments with diverse client data distributions. To prevent overfitting, we implemented an early stopping mechanism, monitoring the evaluation loss with a patience of 10 epochs. This ensured that training would halt if there was no improvement in the loss, thereby maintaining model efficiency and avoiding unnecessary training cycles.

In addition to standard, clean training, we introduced data poisoning as part of our experiments. Specifically, one of the clients (client 3) was designated as the poisoned node across experiments involving 3, 4, and 5 clients. The poisoning attack was executed by selectively altering the labels in client 3's local dataset. To simulate a realistic attack scenario, 70\% of the samples from 6 of the 11 available labels were randomly reassigned to incorrect labels within the dataset, excluding their original labels.

The corrupted data was then used to train the local model on client 3, which introduced malicious gradients during the global model update process. As more rounds of training progressed, the cumulative effect of the poisoned data from client 3 distorted the global model's decision boundary, ultimately reducing the model's overall accuracy and compromising its ability to correctly classify data.

\begin{figure*}[!ht]
\centering
\includegraphics[width=0.95\textwidth]{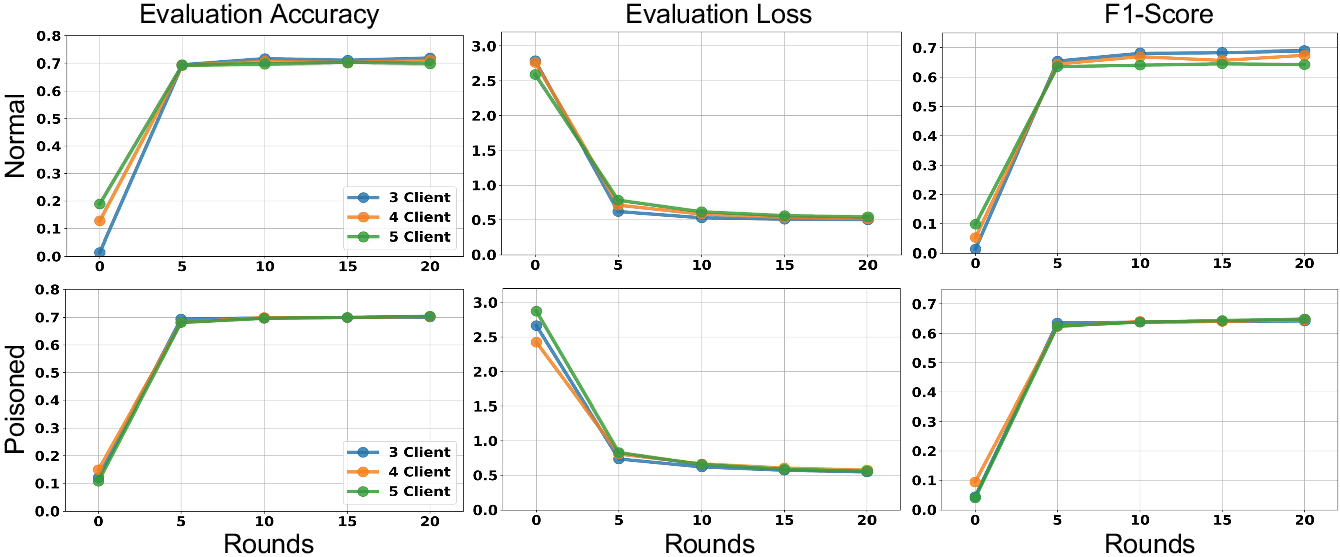}
\vspace{-2mm}
\caption{The performance of the aggregated model of each experiment with various number of clients in evaluation accuracy, loss, and F1-Score}
\vspace{-4mm}
\label{fig:agregatedmodels}
\end{figure*} 

The poisoning attack not only impacted client 3's local model performance but also degraded the performance of the global model shared among all clients. This demonstrates the cascading effect of data poisoning in a FL system, where adversarial manipulation at the local level can propagate through the system, corrupting the global model. By experimenting with different configurations of clients (3, 4, and 5 clients), we were able to observe how the severity of the poisoning varied based on the number of benign clients and the scale of the poisoned data. These experiments provided valuable insights into the vulnerabilities of FL systems under adversarial conditions and how such attacks compromise the integrity and performance of the global model.

\subsection{Clean Performance Results}
In Figure~\ref{fig:normal-5client}, we show the performance of a FL system using non-poisoned data across five clients, with evaluation accuracy, loss, and F1-score tracked over 20 training rounds. Evaluation accuracy steadily improves, converging around 0.70 for all clients, demonstrating consistent learning. The evaluation loss decreases smoothly from 0.8 to below 0.6, indicating effective convergence without disruptions. The F1-score follows a similar pattern, increasing to 0.63–0.67 across clients, reflecting balanced precision and recall. Overall, the system performs robustly under normal conditions, with minimal variation among clients and a successful convergence of the global model.

\vspace{-1mm}
\subsection{Poisoned Model Performance Results}
Figure~\ref{fig:Client-Poisoned} presents the performance of each client across three experiments with different numbers of clients (3, 4, and 5) in FL settings. The evaluation metrics displayed include evaluation accuracy, evaluation loss, and F1-score over 20 rounds of training. Each experiment involves a unique configuration of clients, with client 3 being consistently poisoned across all trials. 

In the case of 3 clients, we observe a significant drop in evaluation accuracy and F1-score for client 3 compared to clients 1 and 2. This degradation indicates that client 3's poisoned data severely impacts its local model's ability to perform accurately. Furthermore, the evaluation loss for client 3 remains higher throughout the rounds, signifying that the poisoned data is contributing to poor model convergence. In the 4-client and 5-client scenarios, the introduction of additional clients helps mitigate the effect of the poisoned data. As more clients are added, the global model appears to be more resilient to the poisoned contributions from client 3, as seen by the relatively stable evaluation accuracy and F1-scores for the non-poisoned clients. However, client 3's performance continues to lag significantly, highlighting the local impact of the poisoning.

As the number of clients increases to 4 and 5, the overall evaluation loss and performance trends suggest that FL becomes more robust to individual client poisoning. Clients 1, 2, 4, and 5 show improvements in both accuracy and F1-score, even with the presence of a poisoned client. This resilience can be attributed to the aggregation of updates from a larger number of benign clients, which helps to dilute the effect of client 3's poisoned updates. The results suggest that while poisoned clients can have a severe impact on local model performance, the collective aggregation of updates in FL helps maintain global model accuracy, especially as the number of clients increases. However, it also emphasizes the importance of incorporating defense mechanisms to detect and mitigate poisoned contributions, particularly in smaller FL environments where the influence of a poisoned client is more pronounced.

\subsection{Comparison of Aggregated Models}

In Figure~\ref{fig:agregatedmodels}, the performance of the FL system is compared under two conditions: a normal scenario (top row) and a poisoned scenario (bottom row). The results are presented across three metrics: evaluation accuracy, evaluation loss, and F1-score, for configurations of 3, 4, and 5 clients. 

In the normal scenario (top row), the evaluation accuracy shows a steady increase during the first 5 rounds, stabilizing around 0.7 for all client configurations (3, 4, and 5 clients). The evaluation loss also significantly decreases over time, with a sharp drop in the first 5 rounds before leveling out as the model converges. The F1-score follows a similar pattern, increasing rapidly in the first few rounds and stabilizing at around 0.7. The similar behavior of the different client configurations in terms of accuracy, loss, and F1-score indicates that in the absence of poisoning, the FL model performs consistently well, regardless of the number of clients involved.

In contrast, the poisoned scenario (bottom row) exhibits a notable degradation in performance, especially in evaluation accuracy and F1-score. While the initial increase in accuracy during the first few rounds appears similar to the normal scenario, the final accuracy levels off at a slightly lower value compared to the normal condition. Evaluation loss also shows a similar decreasing trend, but it does not reach as low a value as in the normal condition, indicating that the model struggles to fully converge when poisoned data is introduced. Additionally, the F1-score, while following the same upward trajectory, stabilizes at a marginally lower value compared to the normal scenario. 

These results show that while FL systems can maintain some level of resilience to poisoned data, the performance in terms of accuracy and F1-score is slightly reduced. The model does not fully converge as effectively as in the normal scenario, highlighting the need for further improvements in defense mechanisms to mitigate the impact of adversarial attacks like data poisoning.

\vspace{-1mm}
\section{Conclusion and Future Work}
\vspace{-1mm}

In this paper, we developed and evaluated a FL testbed to assess the resilience of models in the face of adversarial data poisoning, specifically in a cybersecurity use case involving DNP3 intrusion detection. Through our experiments with both normal and poisoned data, we demonstrated that while FL systems show robust performance under normal conditions, poisoned data significantly impacts local model accuracy and convergence. However, as the number of clients increases, the global model remains more resilient due to the aggregation of updates from non-poisoned clients. The use of the Flower framework proved highly effective for its flexibility, scalability, and compatibility with resource-constrained devices like the Nvidia Jetson Nano. Our findings underscore the need for robust defense mechanisms to address FL's susceptibility to adversarial attacks in critical applications like cybersecurity. 

Future work includes implementing and evaluating specific defenses, such as various Byzantine Robust Aggregation algorithms (e.g., distance-based Krum, statistics-based Median) discussed in \cite{Cajaraville-Aboy_2024}. We also plan to integrate differential privacy by adding calibrated noise to client updates before aggregation, assessing the privacy-utility trade-off as analyzed in other frameworks \cite{Wei_2019}. Additionally, we plan to explore more complex attack strategies to further stress-test FL models and inform the development of secure and resilient FL systems for privacy-sensitive environments.

\vspace{-2mm}
\bibliographystyle{IEEEtran}
\bibliography{ref}

\end{document}